\documentstyle[twoside,fleqn,espcrc2,epsf]{article}

\def\CO{{\cal O}}
\def\CL{{\cal L}}

\def\semitimes{\mathrel>\joinrel\mathrel\triangleleft}

\title{
\vspace{-0.5cm}
Partial Restoration of Flavor Symmetry for Chiral Staggered Fermions
\thanks{Supported by DOE contracts DE-FG03-96ER40956 
and DOE-W7405-ENG-86.}
}

\author{
Weonjong Lee\address{Los Alamos National Lab, MS B-285, Los Alamos,
		New Mexico 87545, USA}
and
Stephen Sharpe\address{Department of Physics, Box 351560,
University of Washington, Seattle, WA 98195, USA}\thanks{Speaker.}
}
       
\begin{document}

\begin{abstract}
At non-zero lattice spacing the flavor symmetry of staggered fermions is
broken to a discrete subgroup. We show that in the chiral limit the
flavor symmetry of the pion effective Lagrangian enlarges to an 
$SO(4)$ subgroup of the continuum $SU(4)$ symmetry.
This provides an explanation for observed degeneracies in the pion spectrum.
\end{abstract}

\maketitle

A single species of staggered fermion represents four degenerate
quarks in the continuum limit, but at finite lattice spacing
the $SU(4)$ flavor symmetry is broken down to a discrete 
subgroup\cite{GS}. We have found, however,
that there is a substantial
restoration of the flavor symmetry in the chiral limit, 
at least in the pion sector \cite{LS}.
We give here a description of our conclusions and an outline of
the arguments supporting them. 
Our analysis is an application of the method used
in \cite{ss} to study flavor breaking with Wilson fermions.

We begin by recalling how flavor and rotation symmetries are
broken by staggered fermions. 
In the continuum limit, the four flavors can be described by
a field $Q_{\alpha,a}$, 
with spinor and flavor indices $\alpha$ and $a$, respectively.
The Euclidean theory has an\footnote{%
Other symmetries of the continuum theory will not be 
important here.
Fermion number and the discrete symmetries (e.g. parity)
are not broken by the discretization,
while translation symmetry is broken in the obvious way.}
\begin{equation}
SU(4)_{\rm flavor} \times SO(4)_{\rm rotation}
\label{eq:contsymm}
\end{equation}
symmetry, under which the quarks transform as
$Q \to (\Lambda_{1/2} \otimes U) Q$, with $\Lambda_{1/2}$ in the
spinor rep. of $SO(4)$ and $U$ an $SU(4)$ matrix.
To obtain the subgroup of (\ref{eq:contsymm}) respected on lattice,
it is useful to consider first an intermediate subgroup,
obtained by restricting flavor $SU(4)$ matrices to be those of the
$SO(4)$ spinor rep:
\begin{equation}
U \longrightarrow \Lambda_{1/2}=\exp(i \omega_{\mu\nu} \sigma_{\mu\nu}) \,.
\end{equation}
In addition, we keep a discrete set of matrices,
$(i\gamma_\mu\gamma_5)$ which are contained in $SU(4)$, but
not in $SO(4)$.
These generate a 4-dimensional 
Clifford algebra, $\Gamma_4$, and rotate as a vector
under $SO(4)$. Thus a subgroup of (\ref{eq:contsymm}) is
\begin{equation}
\Gamma_4 \semitimes SO(4)_{\rm flavor} \times SO(4)_{\rm rotation}
\label{eq:restsymm}
\end{equation}
($\semitimes$ indicates a semidirect product).

Finally, consider the hypercubic subgroups of the two $SO(4)$ groups,
which we refer to as $SW_4$ \cite{toolkit}.
In particular, let $SW_{4,\rm diag}$ be the hypercubic group in which
flavor and spatial rotations are performed simultaneously.
Then the flavor-rotation symmetry of staggered fermions is
\begin{equation}
\Gamma_4 \semitimes SW_{4,\rm diag} \,.
\label{eq:lattsymm}
\end{equation}
On the lattice the $\Gamma_4$ is generated by single-site translations,
while the $SW_{4,\rm diag}$ is composed of rotations.
The details, which are not needed here, can be found in \cite{GS,toolkit}.

We can now state our result: in the chiral limit, at a fixed, but small,
lattice spacing $a$, the effective
flavor-rotation symmetry enlarges from the discrete group (\ref{eq:lattsymm})
to the continuous group (\ref{eq:restsymm}). We have demonstrated that
this restoration holds for the properties of pions, but suspect
that it holds more generally.

Let us illustrate the consequences of this result for the pion spectrum.
In the continuum limit, the pions lie in a 15-dimensional representation of
$SU(4)$, created by the operators
\begin{equation}
\bar Q (\gamma_5 \otimes T^a) Q\,, \quad a=1\!-\!15 \,.
\end{equation}
A convenient choice of flavor matrices is
\begin{equation}
T^a = \left\{ \xi_\mu, (i/2) [\xi_\mu,\xi_\nu], i \xi_\mu\xi_5, \xi_5 \right\}
\,,\quad \xi_\mu=\gamma^*_\mu \,,
\end{equation}
and we use these (dropping factors of $i$ and $i/2$)
to label the states.
On the lattice this 15-plet breaks down into 7 irreps of the (discrete)
symmetry group of the transfer matrix 
[a subgroup of the full lattice group (\ref{eq:lattsymm})]. 
There are \cite{goltmesons}
\begin{eqnarray*}
{\rm three\ 1\!-\!d\ irreps:} &&\! \xi_5,\ \xi_4\xi_5,\ \xi_4,\ {\rm and} \\
{\rm four\ 3\!-\!d\ irreps:} &&\!
 \xi_i\xi_5,\ [\xi_i,\xi_j],\ [\xi_4,\xi_j],\ \xi_i,
\end{eqnarray*}
where $i,j=1\!-\!3$.
At non-zero lattice spacing one expects, in general, that the masses of
pions in the 7 irreps to be different.
Our prediction is that, in the chiral limit, at non-zero but small $a$,
the 7 irreps should collapse into the 4 irreps of flavor $SO(4)$, 
namely those with flavor
\begin{equation}
\xi_5,\ \xi_\mu\xi_5,\ [\xi_\mu,\xi_\nu],\ \xi_\mu\,.
\end{equation}
There are thus three predicted degeneracies 
\begin{eqnarray}
M(\xi_4) = M(\xi_i),&&
M(\xi_4\xi_5) = M(\xi_i\xi_5),\nonumber\\
M([\xi_4,\xi_j]) &=& M([\xi_i,\xi_j]).
\end{eqnarray}

In more detail, the prediction is as follows.
Assume that both $a^2 \Lambda^2$ and $m_q/\Lambda$ are small parameters
($\Lambda$ is an abbreviation for $\Lambda_{\rm QCD}$),
i.e. work close to the continuum and chiral limits.
The general expansion of the pion masses is
\begin{equation}
{M_\pi^2({\rm latt})\over \Lambda^2} \sim 
a^2\Lambda^2 + {m_q \over\Lambda} +
{m_q \over\Lambda} a^2\Lambda^2 + \dots \,.
\label{eq:Mpilat2}
\end{equation}
The ${m_q/\Lambda}$ term is the leading continuum contribution,
and respects the $SU(4)$ symmetry. 
The leading discretization errors are of two types:
(1) The $a^2\Lambda^2$ term, which does not vanish in the chiral limit,
and arises because the $SU(4)$ chiral symmetry is broken on the lattice
even when $m_q=0$;
and (2) The $a^2 m_q \Lambda$ term, which vanishes in the chiral limit.
Our result is that the former term respects the intermediate
symmetry group (\ref{eq:restsymm}), while the latter breaks the symmetry
completely down to the lattice group (\ref{eq:lattsymm}).
Thus in the chiral limit, for fixed $a$, the spectrum has an $SO(4)$
symmetry. Note that in this limit there is also an exact 
axial $U(1)$ symmetry, and so the $a^2 \Lambda^2$ term is
absent for the corresponding pion (which has flavor $\xi_5$).

\begin{figure}[t]
\vskip0.1cm
\epsfxsize=3.0in
\centerline{\epsffile{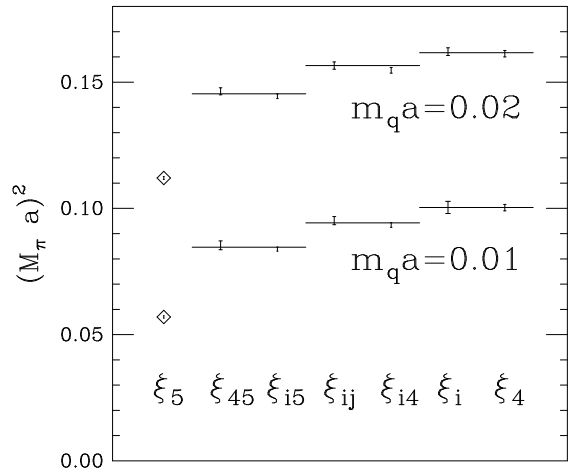}}
\vspace{-0.6truein}
\caption{
$M_\pi^2 a^2$ at $\beta=6$ from Ref. \cite{ishizuka}.
Flavor is listed along the bottom, with the pseudo-Goldstone pion
at the left. Horizontal lines show
the average value of the mass-squared of the pairs of states
which are predicted to become degenerate in the chiral limit.
}
\vspace{-0.3truein}
\label{fig:ishi}
\end{figure}

How well does this prediction work?
The complete spectrum of pions was first calculated
with sufficient precision by JLQCD\cite{ishizuka},
and has recently been obtained for both unimproved and improved
staggered fermions by Orginos and Toussaint \cite{orginos}.
We plot the former results in Fig.~\ref{fig:ishi}. The quark masses
correspond to roughly $1/3$ and $2/3$ of the strange quark mass.
The predicted pattern is seen to hold:
in most cases, the
differences between the non-Goldstone pions in different $SO(4)$
irreps are significant,
while pions in the same $SO(4)$ irrep have equal masses within errors.
This pattern has apparently not previously been noted.
The predictions are also well verified by the results of \cite{orginos}.
Apparently, one does not need to use extremely small quark masses
in order to see the symmetry restoration.

We close this talk with an outline of our argument. As in \cite{ss}
we proceed in two steps: (1) construct an effective continuum Lagrangian,
$\CL_{\rm eff}$ (in terms of quarks and gluons)
describing the long-distance modes of the lattice theory; and
(2) construct an effective chiral Lagrangian 
(in terms of mesons and baryons) describing the long-distance physics
of ${\cal L}_{\rm eff}$.
Following \cite{symanzik}, $\CL_{\rm eff}$ contains all operators
allowed by the symmetries of the underlying lattice theory.
It takes the form
\begin{equation}
\CL_{\rm eff} = \CL_{\rm QCD} + a^2(\CL_{\rm glue}^{(6)}
+\CL_{\rm bilin}^{(6)} + \CL_{\rm ff}^{(6)})+ O(a^4)\,,
\end{equation}
where the superscript indicates the dimension of the operators.
The three dimension-6 terms contain gluonic, quark bilinear and four-fermionic
operators, respectively. 
There are no fermionic operators of dimension 5 because
of the symmetries of staggered fermions \cite{sslat93,luo1}.

We are interested in the flavor and rotation symmetry breaking caused
by the $a^2$ terms. The gluonic operators are flavor singlets, but 
break the rotation symmetry down to the hypercubic group: $SO(4)\to SW_4$.
The fermion bilinears, which have been listed by Luo \cite{luo2},
turn out also not to break flavor.
Thus flavor symmetry breaking is caused entirely by the four-fermion
operators. Luo found 18 such operators; we have found 6 more.\footnote{%
Some of these 24 operators are redundant, but this overcounting takes care of 
itself automatically when we calculate physical quantities 
such as pion masses.}
These 24 operators divide into two types:
(A) Lorentz singlets, which break flavor $SU(4)\to SO(4)$, e.g.
\[
\CO_1 =
\sum_{\mu\nu} \left\{[\bar Q(\gamma_\mu\otimes\xi_\nu) Q]^2
- [\bar Q(\gamma_\mu\gamma_5\otimes\xi_\nu\xi_5) Q]^2 \right\}
\]
(B) Operators which break the 
symmetry down to the discrete
lattice subgroup (\ref{eq:lattsymm}), e.g.
\[
\CO_2 =
\sum_{\mu} \left\{ [\bar Q(\gamma_\mu\otimes\xi_\mu) Q]^2 
- [\bar Q(\gamma_\mu\gamma_5\otimes\xi_\mu\xi_5) Q]^2 \right\}
\]
The key point is that $\CO_2$ is not invariant under separate spatial
and flavor rotations. In fact, one can show that all type (B) operators 
can be chosen to be non-singlets under spatial rotations.
In the present case $\CO_2-\CO_1/4$ is such a non-singlet.

Now we proceed to the second step, and map $\CL_{\rm eff}$ into
the chiral Lagrangian. The dimension-4 operators map as usual:
\begin{equation}
\CL_{\rm QCD} \to
\Lambda^2 {\sf Tr}(\partial_\mu \Sigma \partial_\mu \Sigma^{\dagger})
+ m\Lambda^3 {\sf Tr}(\Sigma+\Sigma^\dagger),
\end{equation}
where $\Sigma=\exp(i\pi^a T^a)$. The flavor singlet $O(a^2)$ terms give
corrections of relative size $a^2 p^2$,
\begin{equation}
a^2 \CL_{\rm glue+bilin} \to
a^2 \Lambda^2 \sum_\mu
{\sf Tr}(\partial_\mu^2 \Sigma \partial_\mu^2\Sigma^{\dagger})
+ \dots \,,
\end{equation}
which break $SO(4)\to SW_4$.
Four-fermion operators of type (A), being Lorentz singlets,
can be mapped into operators in the chiral Lagrangian without derivatives,
e.g.
\begin{equation}
\CO_1 \to \sum_\nu \left\{
{\sf Tr}(\Sigma \xi_\nu\Sigma^\dagger \xi_\nu) -
{\sf Tr}(\Sigma \xi_\nu\xi_5\Sigma^\dagger \xi_5\xi_\nu) \right\}.
\end{equation}
They thus contribute to the $a^2\Lambda^2$ term in (\ref{eq:Mpilat2}),
and break $SU(4)\to SO(4)$.
The key point, however, is that, since
type (B) operators are Lorentz non-singlets, 
their chiral representatives must contain extra derivatives, e.g.
\begin{eqnarray}
\lefteqn{
\CO_2-\CO_1/4 \to \sum_\mu\left\{
{\sf Tr}(\partial_\mu\Sigma \xi_\mu\partial_\mu\Sigma^\dagger \xi_\mu) \right.
}
\nonumber\\ 
&&\mbox{} \qquad\qquad \left. -{\sf Tr}(\partial_\mu \Sigma \xi_\mu\xi_5
\partial_\mu\Sigma^\dagger \xi_5\xi_\mu) \right\}.
\end{eqnarray}
Since $p^2\to M_\pi^2 \sim m_q\Lambda$ on-shell,
these operators only
contribute to the $a^2 m_q\Lambda$ term in (\ref{eq:Mpilat2}).
They break the continuum symmetry down
to the lattice group (\ref{eq:lattsymm}).

We conclude with some general comments. 
Our predicted symmetry restoration applies also for perturbatively
improved staggered fermions, such as those studied in \cite{orginos}.
We stress, however, that we make no predictions for the values of
the splittings between $SO(4)$ multiplets---these will
depend on the discretization. Finally,
as far as we can see, similar symmetry restoration
should hold for other hadron masses.


\begin{thebibliography}{9}

\bibitem{GS} M. Golterman and J. Smit, Nucl. Phys. B245 (1984) 61,
B255 (1985) 328.

\bibitem{LS} W. Lee and S. Sharpe, in preparation.

\bibitem{ss} S. Sharpe and R. Singleton Jr., 
{Phys. Rev.} {D58}, 074501 (1998).

\bibitem{toolkit} G. Kilcup and S. Sharpe, Nucl. Phys. {B283} (1987) 493.

\bibitem{goltmesons} M. Golterman, Nucl. Phys. {B273} (1986) 663.

\bibitem{ishizuka} N.Ishizuka {\em et al.}, Nucl.Phys. {B411}(1994)875.

\bibitem{orginos} K.Orginos and D.Toussaint, hep-lat/9805009, and
these proceedings.

\bibitem{symanzik} K.Symanzik,
%
Nucl.Phys.{\bf B226}(1983)187,205.

\bibitem{sslat93} S.~Sharpe, Nucl. Phys. B(Proc.Suppl.) 34 (1994) 403.

\bibitem{luo1} Y. Luo, {Phys. Rev. D55} (1997) 353.

\bibitem{luo2} Y. Luo, {Phys. Rev. D57} (1998) 265.
\end{thebibliography}
\end{document}